%% file: MAIN.tex
\documentclass[manuscript,sigconf]{acmart}
\AtBeginDocument{%
  \providecommand\BibTeX{{%
    \normalfont B\kern-0.5em{\scshape i\kern-0.25em b}\kern-0.8em\TeX}}}

\copyrightyear{2024}
\acmYear{2024}
\setcopyright{rightsretained}
\acmConference[CHI '24]{Proceedings of the CHI Conference on Human Factors in Computing Systems}{May 11--16, 2024}{Honolulu, HI, USA}
\acmBooktitle{Proceedings of the CHI Conference on Human Factors in Computing Systems (CHI '24), May 11--16, 2024, Honolulu, HI, USA}
\acmDOI{10.1145/3613904.3642427}
\acmISBN{979-8-4007-0330-0/24/05}

\usepackage{soul}

\usepackage{cleveref}
\usepackage{subcaption}
\usepackage{hyperref}
\usepackage{graphicx}
\usepackage{xcolor}
\definecolor{lightBlue}{HTML}{4DBEEE}
\definecolor{darkBlue}{HTML}{0072BD}
\definecolor{brightYellow}{HTML}{EDB120}
\definecolor{purple}{HTML}{7E2F8E}
\definecolor{orange}{HTML}{D95319}
\definecolor{green}{HTML}{77AC30}

\newcommand{\mysym}[2][black]{\textcolor{#1}{\ding{#2}}}
\usepackage{multirow}
\usepackage{tabularx}
\newcolumntype{b}{X}
\newcolumntype{s}{>{\hsize=.08\hsize}X}
\usepackage{pifont}

\renewenvironment{quote}
  {\list{}{\rightmargin=.9cm \leftmargin=.9cm}%
   \item\relax}
  {\endlist}

\usepackage{pgf, tikz}
\usetikzlibrary{arrows, automata}
\begin{document}



\title[Understanding Entrainment in Human Groups]{Understanding Entrainment in Human Groups: Optimising Human-Robot Collaboration from Lessons Learned during Human-Human Collaboration}

\author{Eike Schneiders}
\email{eike.schneiders@nottingham.ac.uk}
\affiliation{%
  \institution{University of Nottingham}
  \streetaddress{Wollaton Road}
  \city{Nottingham}
  \country{UK}
  \postcode{NG8 1BB}
}

\author{Christopher Fourie}
\email{ckfourie@mit.edu}
\affiliation{%
  \institution{Massachusetts Institute of Technology}
  \streetaddress{}
  \city{Cambridge, MA}
  \country{USA}
  \postcode{}
}

\author{Stanley Celestin}
\email{sc3246@cornell.edu}
\affiliation{%
  \institution{Cornell University}
  \streetaddress{107 Hoy Road}
  \city{Ithaca, NY}
  \country{USA}
  \postcode{14853}
}


\author{Julie Shah}
\email{julie_a_shah@csail.mit.edu}
\affiliation{%
  \institution{Massachusetts Institute of Technology}
  \streetaddress{}
  \city{Cambridge, MA}
  \country{USA}
  \postcode{}
}

\author{Malte Jung}
\email{mfj28@cornell.edu}
\affiliation{%
  \institution{Cornell University}
  \streetaddress{107 Hoy Road}
  \city{Ithaca, NY}
  \country{USA}
  \postcode{14853}
}

\renewcommand{\shortauthors}{Schneiders et al.}

\begin{abstract}
  Successful entrainment during collaboration positively affects trust, willingness to collaborate, and likeability towards collaborators. In this paper, we present a mixed-method study to investigate characteristics of successful entrainment leading to pair and group-based synchronisation. Drawing inspiration from industrial settings, we designed a fast-paced, short-cycle repetitive task. Using motion tracking, we investigated entrainment in both dyadic and triadic task completion. Furthermore, we utilise audio-video recordings and semi-structured interviews to contextualise participants' experiences. This paper contributes to the Human-Computer/Robot Interaction (HCI/HRI) literature using a human-centred approach to identify characteristics of entrainment during pair- and group-based collaboration. We present five characteristics related to successful entrainment. These are related to the occurrence of entrainment, leader-follower patterns, interpersonal communication, the importance of the point-of-assembly, and the value of acoustic feedback. Finally, we present three design considerations for future research and design on collaboration with robots.
\end{abstract}



\begin{teaserfigure}
    \centering
    \includegraphics[trim={0cm 0cm 0cm .1cm},clip,width=\textwidth]{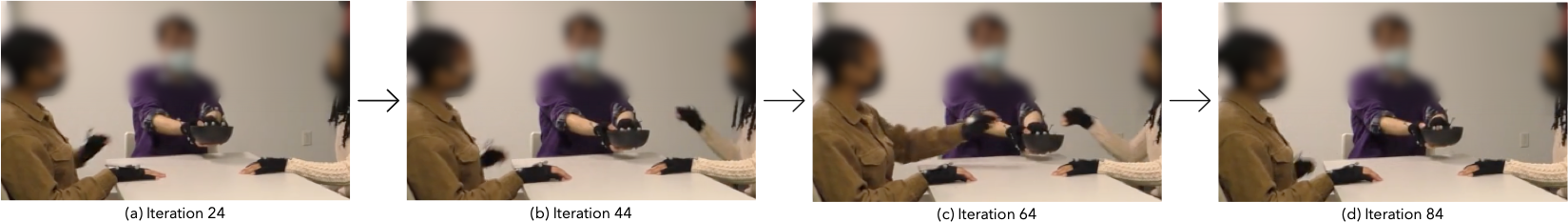}
    \caption{The four frames (left to right) show a snapshot of the triadic collaboration (T4) from iteration 24, 44, 64, and 84. It shows the consistency of the position of the bowl, i.e., the point of assembly, across these 60 iterations.} 
    \label{fig:ConsistentPlacement}
\end{teaserfigure}

\maketitle

\input{Chapters/Introduction}
\input{Chapters/RelatedWork}

\input{Chapters/Study}
\input{Chapters/Results}

\input{Chapters/Discussion}
\input{Chapters/Conclusion}

\begin{acks}
We would like to express our gratitude to the study participants for their participation. Furthermore, we would like to express our thanks to the additional participants who helped with the pilot testing of the tasks. This work was supported by the Engineering and Physical Sciences Research Council [grant number EP/V00784X/1] UKRI Trustworthy Autonomous Systems Hub and the Office of Naval Research [N00014-19-1-2299]. Any opinions, findings, and conclusions expressed in this material are those of the author(s).
\end{acks}

\bibliographystyle{ACM-Reference-Format}
\bibliography{biblio}

\end{document}

%% file: Chapters/Introduction.tex
\section{Introduction}\label{sec:Introduction}

The industrial sector remains one of the fastest growing application areas for collaborative robotics~\cite{Sanneman:2021}.
No other domain has benefited to the same extent from the introduction of robotic automation as the industrial sector has. Given the complexity of modern manufacturing processes, these tasks often require collaboration between multiple, human and non-human, actors. Yet, most studies within the field of Human-Robot Interaction (HRI) emphasise the investigation of dyadic interaction~\cite{Schneiders:2022:THRI}, i.e., the investigation of the interaction between \textit{one} human and \textit{one} robot. However, group-based collaboration and entrainment has received little attention.  
We argue, that efficient collaboration around a given tasks becomes even more relevant given complex group configurations. Prior research has highlighted the importance of temporal synchronisation between collaborators, i.e., rhythms between interaction partners, for efficient collaboration~\cite{Park:2013:ExerSync,Robinson:2020:Boat,Tarr:2018:Synchrony}. To achieve this temporal synchronisation, it is vital that collaborators entrain with one another, thereby achieving better coordination. As Cross et al.~\cite{Cross:2019} state: \textit{``Entrainment refers to temporally coupled or synchronised systems, and it is the process of things moving in time together''}. In other words, entrainment refers to the process of moving in temporal synchronisation. 
In this paper, the term `system' refers to the collaborating partners. Furthermore, while multiple types of entrainment exist, this study focuses exclusively on physical entrainment between different actors, referred to as interpersonal motor synchronisation (IMS)~\cite{Rinott:2021}, thereby excluding other forms of entrainment, such as lexical entrainment.

Achieving temporal synchronisation during collaboration through the process of entrainment leads to a multitude of benefits: it fosters a stronger sense of togetherness and connection~\cite{Hove:2009}, enhances the likeability between collaborators~\cite{Hove:2009,Launay:2014}, and promotes a willingness to cooperate~\cite{Wiltermuth:2009,Rinott:2021}. Entrainment, its occurrence, and its effects on human collaborators have been explored in various contexts, including dancing~\cite{Tarr:2016,Woolhouse:2016}, singing~\cite{Anshel:1988,VonZimmermann:2016}, walking~\cite{vanUlzen:2008}, and body movement~\cite{Tarr:2015,Reddish:2013}. 
The focus on collaborative automation, with an emphasis on human-robot collaboration (HRC) over full automation, has recently received more attention~\cite{Indu5,Indu5:EU}. To fully utilise the strengths of both human and automation technology, such as robots, a deeper understanding of how humans and robots can collaborate efficiently is essential. In our study, we adopt a human-centred approach. Specifically, we investigate human entrainment in dyads and triads to understand the entrainment process among humans, thereby informing design considerations for future research and design of collaborative robots for human-robot collaboration. Enhancing our understanding of how humans achieve collaborative rhythms will pave the way for more effective and natural collaborations within mixed human-robot teams.




In this paper, we investigate how human dyads and triads entrain with each other to improve collaboration during the completion of a fast-paced, short-cycle repetitive task. To investigate this, we conducted a controlled mixed-method laboratory study in which ten dyads and ten triads completed a collaborative task inspired by industrial work. 
We collected a variety of data streams, including motion tracking data, video recordings, and semi-structured post-task interviews. Using video inspection, thematic analysis, as well as trajectory analysis of the motion tracking data, we identified when collaborators achieved a collaborative rhythm and what led to this. 
Lastly, we discuss the implications of the presented findings and propose three design considerations for future Human-Robot Collaboration design and research, increasing the potential for efficient entrainment and ultimately improving human-robot collaboration. 

%% file: Chapters/RelatedWork.tex
\section{Related Work}

This section will present existing research on entrainment during human collaboration, benefits of temporally synchronised collaboration, and physical entrainment in human-robot collaboration.



\subsection{Entrainment in Human Collaboration}\label{sec:RW_HH}
The theory of joint action, a field of cognitive psychology, studies how human agents collaborate in dyadic and group interactions. Entrainment is highlighted as a core process, often described as an unconscious synchronisation between agents across various contexts. Previous research has investigated how pairs and groups of humans physically entrain with each other during task completion and what effects this has on interpersonal relations. Examples include walking in groups~\cite{Wiltermuth:2009}, finger tapping to an external stimuli~\cite{Hove:2009}, observation of animated movement synchronicity~\cite{Lakens:2010}, collaborative biking~\cite{Rinott:2021}, hand-over tasks~\cite{Roy:2020}, as well as literature reviews investigating several aspects of human-human entrainment~\cite{Cross:2019,Rinott:2021}. 
Through an extensive literature review, Cross et al.~\cite{Cross:2019} identify 48 studies focusing on the effects of pro-group behaviour caused by interpersonal entrainment. Various experimental tasks (e.g., dancing, tapping, arm movement) and experiment durations (from 11 seconds to 1 hour) have been conducted to investigate interpersonal entrainment. As summarised by Cross et al.~\cite{Cross:2019}, the investigation of entrainment has received extensive focus in interactions such as body limb movement~\cite{Tarr:2015,Reddish:2013}, dancing in groups~\cite{Tarr:2016,Woolhouse:2016} or singing~\cite{Anshel:1988,VonZimmermann:2016}.

\citet{Roy:2020} conducted a lab-based study to investigate how pairs of human shelf fillers entrain with each other. This study was inspired by the working context observed at supermarkets. They conducted a mixed-method study utilising field studies to better understand supermarket employees' tasks and re-create those in a controlled lab environment. Using both subjective data, such as a verbal description of the experience during task completion or post-session questionnaires, as well as quantitative performance metrics, including `the number of bottles for each 10-second interval' or `level of coordination amongst team members', Roy and Edan identified several factors related to dyadic entrainment in short-cycle repetitive tasks. They observed different behaviour, as shelving bottles included two clear roles (giver and receiver). For instance, while the giver uses visual cues to move the bottles towards the receiver, the receiver does not often look away from the shelf, simply assuming that the point-of-handover will be the same for every shelving cycle. Furthermore, the authors could observe a clear improvement over time in both the consistency of working speed and coordination, which indicates a better synchronisation achieved through entrainment. While Roy and Edan's study investigated dyadic entrainment in one particular setting (bottle shelving), these lessons might be transferable to other tasks and group formations.

\citet{Rinott:2021} present a literature review on interpersonal motor synchronisation-based literature. They propose a framework mapping out different types of Joint-Action---which does not require any type of temporal synchronisation as long as all actors work towards the same goal. As part of this framework, the authors classify `Synchronisation' as the result of `Entrainment': \textit{``Entrainment is a process of reaching the same rhythm and is a required part of synchronising with someone else.''}~\cite[Section 2.1]{Rinott:2021}. Furthermore, based on existing literature, they synthesise eleven dimensions (e.g., temporality, information exchange, or the number of participants) relevant to the study of IMS. Specifically, an important aspect during the investigation of IMS is the temporal aspect. IMS can be investigated using external entrainment, such as a metronome (e.g.,~\cite{Hove:2009}) providing a rhythm, or using `Mutual entrainment', in which participants entrain with each other. Alternative names for the same phenomena have been used, including `Social entrainment' and `Mutual social entrainment'~\cite{Phillips-Silver:2010}.

Previous research has investigated a variety of tasks---such as shelf filling, walking, or dancing---and how human dyads and groups entrain in these. We argue that studying tasks resembling industrial context is important to understand entrainment in this specific context. Furthermore, concrete and context-specific design considerations can be presented by relating the study of group-based entrainment to one specific context, thereby optimising human-robot collaboration for industrial tasks. 



\subsection{Temporally Synchronised Collaboration}
Prior research has highlighted positive effects achieved through successful temporal synchronisation, as a direct result of entrainment~\cite{Rinott:2021}, amongst human collaborators. Examples include an increased sense of togetherness~\cite{Hove:2009}, optimised task performance~\cite{Valdesolo:2010}, higher likeability amongst collaborators~\cite{Hove:2009,Launay:2014}, willingness to cooperate~\cite{Wiltermuth:2009,Rinott:2021}, trust~\cite{Rinott:2021}, the sense of connectedness~\cite{Wiltermuth:2009,Rinott:2021}, and rapport~\cite{Lakens:2010,Miles:2009,Lakens:2011:MoveInSynchRapport}.

The desire for increased efficiency during task completion is one of the primary motivations for collaboration. While this can be achieved through temporal synchronisation~\cite{Valdesolo:2010}, efficient collaboration benefits from other aspects such as a higher willingness to cooperate (e.g.,~\cite{Wiltermuth:2009,Rinott:2021}). To identify ways to prevent the free-rider problem during group collaboration, i.e., members of the group who do not contribute a fair share to the collective effort, Wiltermuth and Heath~\cite{Wiltermuth:2009} have investigated the relationship between the willingness to cooperate and temporal synchronisation. The authors completed three studies in which participants i) walked in synchrony, ii) were addressed as a group instead of individuals, and iii) moved in synchrony. Following each activity, participants cooperated during experimental tasks. Results showed that in all three studies, the participants who, prior to the experimental task, were in the synchrony condition performed better as a group in contrast to the control group. Thereby, Wilterumth and Heath showed a clear indication of the beneficial effect on the willingness to cooperate, leading to better results when acting in synchrony.

Numerous studies have shown that synchronisation with interaction partners increases the sentiment toward each other. For example, Valdesolo et al.~\cite{Valdesolo:2010} investigated if being synchronised not just influences perception but also \textit{improves} performance on specific tasks. Participants completed a joint action coordination task, requiring the anticipation of the collaborator's movement, after rocking on rocking chairs either synchronously or asynchronously. Following this, participants collaboratively navigated a steel ball through a labyrinth. Results showed that pairs rocking in synchrony had a significantly higher sense of similarity and connectedness but were also more efficient at task completion.

Launey et al.~\cite{Launay:2014} present a study in which participants tap to the beat of sounds, either synchronously or asynchronously. Even though the partner was not physically present but shown as a video recording, participants had a higher degree of likeability towards the `virtual partner' when tapping in synchronisation compared to tapping asynchronous. Interestingly, the study showed that when participants believed that a collaborator did not make the tapping sound but by a computer, no difference in likeability was detected depending on if the tapping was synchronous or asynchronous.





\subsection{Physical Entrainment in HRC}
The occurrence of entrainment has not just been observed in human-human teams. Prior research has investigated if entrainment arises during collaborative tasks with robots (e.g.,~\cite{Breazeal:2002,Lorenz:2011,Iqbal:2016,Heijnen:2019,Ansermin:2017:Entrainment,Rea:2019:Entrainment,Iqbal:2022:Movement}).

A study by Lorenz et al.~\cite{Lorenz:2011} investigated if the entrainment occurring between dyads of humans also is observable in dyads of humans and robots. They conducted two studies, one investigating human-human entrainment in a simple `reach-forth-and-back' task with human dyads, and a follow-up study, producing similar results, in which one participant per dyad was replaced with a robot. This suggests that cooperative task completion can result in entrainment, even if one of the collaborators is a robot. These findings are particularly interesting, given the movement speed of the robot. In the human-human condition, participants entrained to one another and reached temporal synchronisation, i.e., both partners reached start and end points simultaneously. However, the robot's movement speed was about a third slower than the average movement speed during the human-human task. Nevertheless, temporal synchronisation between humans and robots occurred, indicating that the human interaction partner reduced their speed to synchronise with the robot.

In line with Lorenz et al.~\cite{Lorenz:2011}, Ansermin et al.~\cite{Ansermin:2017:Entrainment} conducted a study to investigate further if entrainment occurs in human-robot dyads. To this end, they conducted an entrainment study using the NAO robot, in which the human interaction partner performs movements in three conditions. During the first condition, three select movements were conducted at the participant's own pace. The second condition added the robot as a movement partner. Here, the NAO would perform the movement while the participants were still instructed to perform them at their own pace. Findings show that all participants were influenced by the robot's movement speed and adapted to match it, leading to unidirectional entrainment. In the third condition, the robot could detect the humans' movement and adapt its movement to it, i.e., bidirectional entrainment. These findings show that human-robot entrainment occurs during body part movement tasks. This was observed for both unidirectional (i.e., human adapting to robot) and bidirectional (i.e., both adapting to each other) entrainment.

In contrast to existing work within human-robot entrainment, our study contributes with a comparative study of pair and group-based (i.e., triads) human-human entrainment, leading to design considerations for future cobot design for improved human-robot collaboration. 

%% file: Chapters/Study.tex
\begin{figure*}[h]
    \centering
    \includegraphics[width=.47\textwidth]{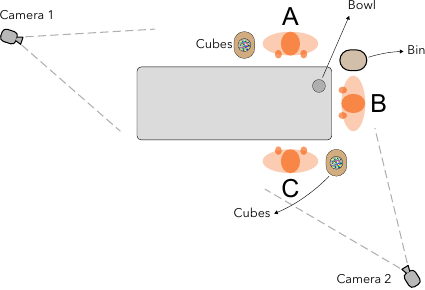}
    \hspace{0.04\textwidth}
    \includegraphics[width=.47\textwidth]{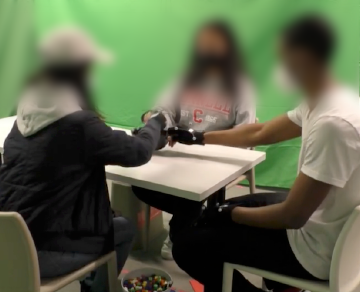}
    \caption{\textit{Triadic setting} - Cubers are placed along the long sides of the table with a bowl of cubes placed next to them. The bowler is placed at the end of the table (position B) with the bowl and the collection bin next to them. The cubers/bowler could choose to place the cubes/bin on their right or left side. The right side image shows the cropped view of Camera 2 (T9).}
    \label{fig:Setup}
\end{figure*}

\section{Methodology} 
This section describes the study as well as the utilised data collection and analysis methods. The conducted study is inspired by previous work on human-human entrainment to inform human-robot entrainment (e.g.,~\cite{Roy:2020,Schneiders:Celestin:JAAE:2022,Schneiders:2023:HHC}).

\subsection{Participants}
We recruited 50 participants (38 female, 12 male, average age: 22.96, SD: 4.06) with varying educational backgrounds, including health, engineering, and art, to mention a few. Participants were recruited using social media postings, flyers on campus, as well as through a dedicated web page. 
The study was approved by the Institutional Review Board for Human Participants at Cornell University (IRB0010723). All participants were compensated with a 50\$ Amazon gift card for their participation. 
We matched participants randomly into ten triads (\textit{N}~=~30, 1--T10) and ten dyads (\textit{N}~=~20, D11--D20). Additional participants were recruited for pilot testing. 

\subsection{The Task}
\subsubsection{Task Development and Pilot testing}\label{sec:taskDev}
To identify suitable tasks in which entrainment has a high potential to naturally occur, we compiled potential ideas resulting in a total of eight different short-cycle repetitive tasks: envelope stamping, tower building, food assembly, block sorting, domino brick placing, packing task, drawing task, and pick-and-placing. After prototyping all eight tasks, we narrowed the selection down to two viable candidates. The final two candidates were piloted both in the dyadic as well as the triadic configuration. Based on our observations as well as the feedback provided by the pilot participants, we selected one final task. The final task (see~\Cref{sec:Task_1}) was chosen as it was easy to learn and required collaboration amongst members of both dyads and triads. Furthermore, the use of a single task was further inspired by existing research~\cite{Roy:2020,Lorenz:2011}. 

\subsubsection{The Task: Pick-and-Place}\label{sec:Task_1}
Inspired by industrial pick-and-placing, we designed a fast-paced repetitive task to investigate entrainment between human collaborators. The goal of the pick-and-placing task was to move small plastic cubes (1$\times$1$\times$1cm) from a bowl (two bowls in the triadic setting) to a collection bin through collaborative effort. To accomplish this task, we defined two distinct roles. The `bowler' and the `cuber(s)'. In the \textit{dyadic setting}, these two roles would be placed across from one another with a table in between them (position A and C in Figure~\ref{fig:Setup} left). It was the cuber's task to pick up \textit{one} cube at a time and drop it in the bowl. When precisely one cube was in the bowl, it was the bowler's responsibility to move the bowl over the collection bin---placed to the right or left of the bowler---and drop the cube from the bowl into the collection bin. Following the emptying of the bowl, the task would repeat. 

In the \textit{triadic setting}, the bowler would sit at the end of the table in position B (Figure~\ref{fig:Setup} left) and the \textit{two cubers} would sit to the right and left of the bowler, respectively (see position A and C in Figure~\ref{fig:Setup} left)\footnote{This submission is accompanied by a video of the task completion by a dyad and a triad. All participants, shown in the video, have given informed consent to the use of the video material.}. In the triadic condition, \textit{exactly} one cube per cuber had to be in the bowl before the bowler could move it to the collection bin. Other than the inclusion of an additional cuber, the task remained the same. For both the dyadic and triadic task completion, the division of roles (who is a cuber and bowler) and the point-of-assembly (i.e., the location where collaborators' actions meet), were decided amongst participants. The point-of-assembly (PoA) is the specific location in which the bowl is positioned, i.e., placed or held, for the cube(s) to be dropped. For an example task completion, please see the video provided in the supplementary material accompanying this paper. The video demonstrates the dyadic and triadic task completion ($\sim$1 minute each).

\begin{table*}[t]
\renewcommand{\arraystretch}{1.2}
\begin{tabularx}{\textwidth}{sb}
\hline
\textbf{Category} & \textbf{Example questions} \\ \hline
\multirow{2}{*}{1} & How would you describe your overall experience with the task? \\
 & If we were to ask you to do it again, what would you do differently? \\ \hline
\multirow{2}{*}{2} & Did you like your role as either bowler or cuber? \\ & Do you think your task was easier or more difficult compared to the other role? \\ \hline
\multirow{2}{*}{3} & During this collaboration you frequently moved the bowl and the cubes. How did you agree on where your individual actions should meet? \\
 & Do you think that your way of coordinating worked well for you? If not – what would you do differently? \\ \hline
\multirow{2}{*}{4} & Did you at any point in the interaction feel uncomfortable? \textit{If so – why?} \\
 & If you had to do this again, would you prefer to work together with a robot or a human next time? \textit{Why?} \\ \hline
\multirow{2}{*}{5} & You just collectively performed several iterations of the task. Did you feel a change in performance? \\
 & Why do you think this change \textit{[if a change was described in the previous questions]} happened? \\ \hline
\end{tabularx}
\caption{Example questions for the post-task semi-structured interviews for each of the five categories. The interviews were conducted as group-based interviews including all collaborators.}
\label{tab:InterviewQues}
\end{table*}

\subsection{Experimental Procedure}
Upon entering the research lab, participants received a participant information sheet  providing them with key information about the study. This included details about its duration and overall structure, purpose of the study, as well as their right to withdraw. This information was then further explained verbally, and participants were given the opportunity to ask any questions they had before proceeding to sign the informed consent form.

The data collection process began with gathering demographic information such as age, gender, and handedness, which was done through a Qualtrics questionnaire. Next, each dyad or triad of participants was introduced to the task, and they were given another chance to ask questions before commencing the task. Participants were instructed to continue the task until interrupted. As our goal was not to assess whether they performed the task slower or faster over time, but rather to investigate fluctuations in its completion consistency, participants were not required to achieve a specific number of cubes. Instead, the emphasis was on completing the task as efficiently as possible, with an undisclosed completion time. After four minutes, we interrupted the participants for the post-task semi-structured interview. 
The interview covered five specific topics: 1) their general experience during task completion, 2) their experience and preference towards the roles they had chosen, 3) their experiences and strategies for negotiation at the point-of-assembly, 4) their level of trust towards the other participants, and 5) their perceived performance. Example questions for each category can be seen in~\Cref{tab:InterviewQues}. As we used a semi-structured interview approach, ad-hoc follow-up questions occurred. The post-task interviews were conducted in dyads or triads to foster insightful conversations about the participants' experiences during the task.




\subsection{Data Collection and Analysis}
For this study, we utilised a mixed-method approach combining qualitative and quantitative measurements. Specifically, we used an OptiTrack camera setup with 18 cameras for motion tracking of participants' hands and the position of the bowl. However, given that the bowl was at all times in the bowlers hand, we did not use its position for the data analysis and visualisation. Additionally, we used two video cameras (see~\Cref{fig:Setup}) to ensure audio-video recordings of participants for later video inspection related to interpersonal communication. Following the task completion, we conducted post-task semi-structured group-based interviews (see~\Cref{tab:InterviewQues} for example questions).


The first author analysed the interview data using thematic analysis in order to identify themes relevant to the investigation of entrainment. Using the motion tracking data, we detected the completion of each iteration of the task. This was done by detecting the point of rotation of the bowlers hand, i.e., the moment the cubes were dropped into the collection bin. Furthermore, the motion tracking data allowed us to visualise dyads and triads performance consistency. This approach allowed us to identify the number of task iterations completed during each 10-second interval~\cite{Rinott:2021}. Furthermore, we used motion-tracking to identify shifts in temporal and spatial variations between task iterations as indicative of successful entrainment. Lastly, we used the video recordings, for visual inspection using ELAN~\cite{Elan}, to contextualise participants' reported experiences.

%% file: Chapters/Results.tex
\section{Results}\label{sec:Results}
We begin by comparing task performance and demographic measures between dyads and triads to establish a suitable basis through which to investigate characteristics of entrainment. Subsequently, this section 
presents five findings related to i) indication of synchronisation, ii) different leader and follower patterns, iii) interpersonal communication, iv) the point-of-assembly, as well as v) the importance of sensory information.


\subsection{Comparability: Dyads and Triads}\label{sec:Performance}
As some of the findings presented are of comparative nature, i.e., highlighting differences in dyadic and triadic task completion, it is important to investigate if dyads and triads are comparable in terms of performance (i.e., completed a similar amount of iterations) and demographic distribution. We started by comparing their respective performance (see~\Cref{tab:DyadTriad}), measured using average iterations per group as well as average iterations per 10 second intervals, as inspired by~\citet{Roy:2020}. \Cref{tab:DyadTriad} shows the comparison of dyadic to triadic performance. For both dyads and triads, an iteration was counted from the time the cube(s) were dropped into the collection bin by the bowler until the next cube was dropped. Two triads where removed from the data analysis due to technical errors during the collection of the motion tracking data.
With this analysis, we initially aimed to examine whether there was a performance difference between dyads and triads.
 

\begin{table}[h]
\renewcommand{\arraystretch}{1.3}
\begin{tabular}{@{}lrr@{}}
\toprule
               & \textbf{Average (SD)} & \textbf{Avg. iterations per 10s (SD)}  \\ \midrule
\textbf{Dyad (\textit{N} = 10)}           &   101.9 (27.847)      & 4.430 (1.211)                                         \\
\textbf{Triad (\textit{N} = 8)}            &   99 (25.890)         & 4.363 (1.084)             \\ \bottomrule
\end{tabular}
\captionsetup{width=.98\linewidth}
\caption{Average performance for the dyads (\textit{N}~=~10) and the triads (\textit{N}~=~8) during task completion. Furthermore, it shows the number of average iterations for each 10 second interval. In none of the two metrics, a significant difference could be identified, which is indicative of the groups being comparable in relation to performance.}
\label{tab:DyadTriad}
\end{table}

Following the performance metrics, we tested whether the distribution of participants into the two conditions resulted in groups with significant differences in relation to demographic representation. Results showed that there were no significant differences among the two groups in terms of gender ($\chi^2$(1, \textit{N}~=~44)~=~0.489, \textit{p}~=~.484), age (\textit{F}(1,42)~=~1.029, \textit{p}~=~.316), or handedness ($\chi^2$(1, \textit{N}~=~44)~=~0.5843, \textit{p}~=~.445).
Lastly, we compared the task performance rate, measured using average completed task iterations pr. 10 second interval~\cite{Roy:2020} as shown in~\Cref{tab:DyadTriad}, between dyadic (\textit{M}~=~4.430, \textit{SD}~=~1.211) and triadic groups (\textit{M}~=~4.363, \textit{SD}~=~1.084). Results of an unpaired t-test showed no significant differences in performance between the two groups, \textit{t}(16)~=~0.124, \textit{p}~=~.903.

Given that no significant differences could be observed, the random distribution of participants into either dyad or triad, or the resulting performance displayed by the groups, we can conclude that the two conditions are indeed comparable.

\subsection{Synchronisation}\label{sec:Synch}

\begin{figure*}[h]
    \centering
    \includegraphics[trim={0cm 0cm 1.2cm 0cm},clip,width=.5\linewidth]{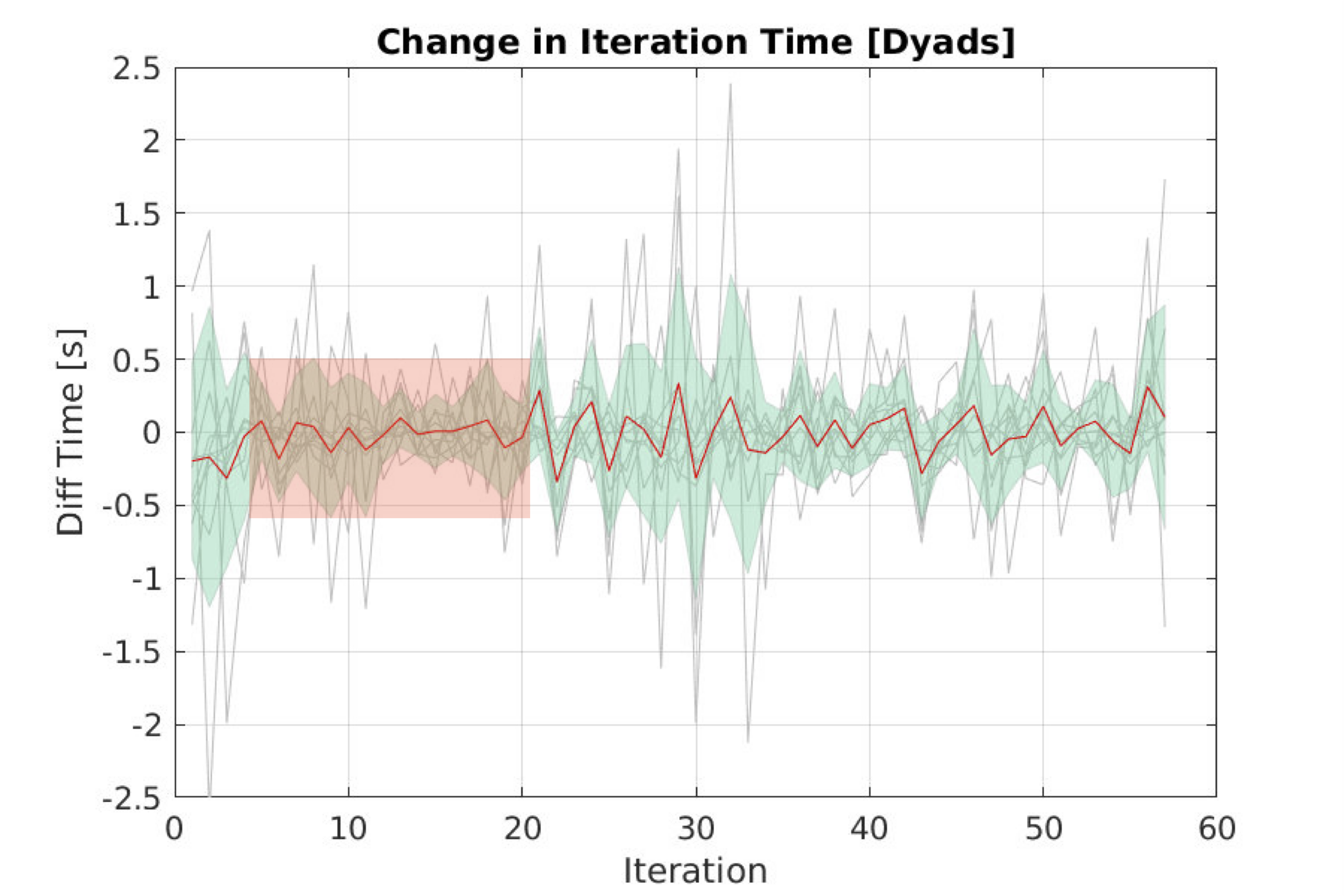}%
    \includegraphics[trim={0cm 0cm 1.2cm 0cm},clip,width=.5\linewidth]{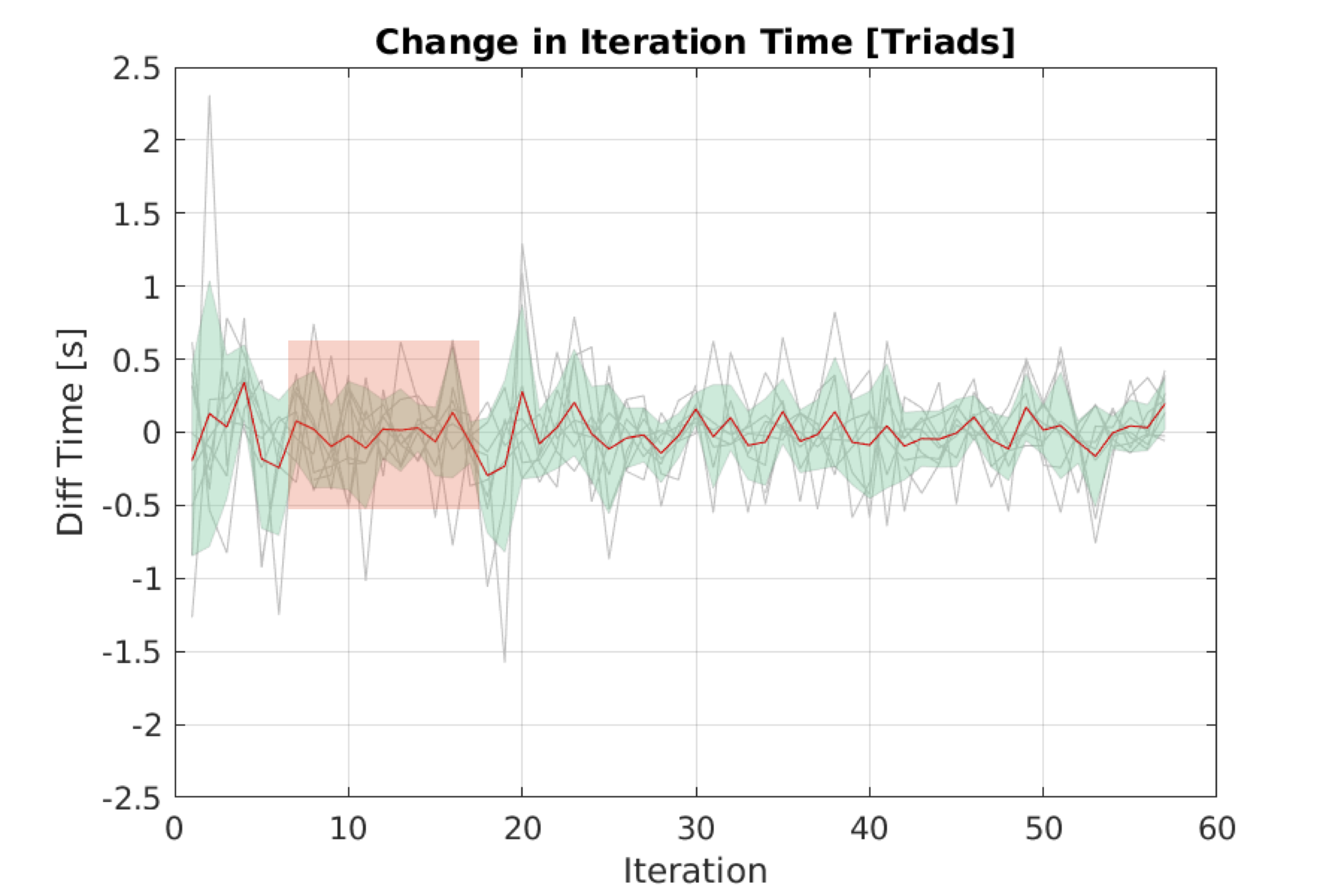}
    \caption{Raw data plotted for dyads (\textit{N}~=~10) and Triads (\textit{N}~=~8) showing the fluctuation times in iteration between each set of consecutive iterations. Furthermore, it plots the average (as a solid red line), the standard deviation (green envelope), as well as the initial period of low temporal fluctuations, suggesting entrainment, period after starting the task (red overlay).}
    \label{fig:Graphs}
\end{figure*}

To investigate the occurrence of entrainment, we analysed the motion tracking data, see~\Cref{fig:Graphs}. Each dyads' and triads' raw data is plotted, with the solid red line representing the average and the green envelope indicating the standard deviation of all dyads (left) and triads (right). The average duration for each task completion of the iteration was 2.26 and 2.29 seconds per iteration for dyads and triads respectively. The graph illustrates fluctuations in task completion time for each iteration, specifically how much faster or slower iteration $i+1$ was compared to $i$, i.e., how fast is the \textit{next} iteration compared to the \textit{last}. A graph with minimal fluctuations, i.e., the time each individual task iteration took is more stable, suggests a higher level of consistency in collaborative rhythm, rather than speed, during task execution. Through manual video inspection, we removed outliers caused by participants dropping cubes, which resulted in extreme iteration times. 

Based on the motion tracking data, we observed multiple periods, of varying duration, of increased consistency, indicated through the reduction in fluctuations in time between iterations. The plotted data led us to hypothesise that both dyads and triads experience an initial period of entrainment within approximately the first five (dyads) to six (triads) iterations\footnote{This initial period of entrainment is highlighted using a red overlay in Figure~\ref{fig:Graphs}.}. This brief duration might be attributed to the task's simplicity, reducing the amount of iterations necessary for successful entrainment. Following this initial entrainment, participants reached a period of consistent time intervals between individual iterations. We observed multiple groups breaking their working rhythm (i.e., temporal fluctuations increase) and attempting to re-optimise their workflow after performing the task for a period of time (i.e., temporal fluctuations decrease).

The phenomenon of synchronisation---also referred to by participants as falling into a `rhythm,' `pattern,' or a `groove,' was not only observed through quantitative measurements but also directly experienced and described by the collaborators (e.g., T1, T4, T6, D11, D13). While the majority of groups noticed that they had fallen into a rhythm, some described it the outcome of an active and conscious effort. For example, T1 actively made an effort to coordinate their movements. They, consciously, attributed the occurrence of synchronisation to this intentional effort. A cuber in T1 for instance expressed that they:
\begin{quote}
    \textit{`[...] in the beginning, we're trying to coordinate and it reaches towards some sort of synchronisation that we [when reached] don't have to pay extra attention as long as everyone's in the same tempo.'} - T1
\end{quote}

A similar observation was made by the bowler in T8, who expressed that \textit{`I think I kind of adjusted to their speed. Because [...] at first I think I was going too fast.'} In the above quotes, collaborators describe the active effort of coordinating to perform the given task efficiently. Through this effort `synchronisation', a `rhythm', or a `groove' is reached, allowing them to complete the task without paying active attention to this coordinating. The participant talked about `synchronisation' as being related to the tempo in which group members perform their individual sub-task. Using the video recordings, we observed that reaching synchronisation allowed participants to start conversing---both on- as well as off-topic (see~\Cref{sec:Communication}).
Once a common rhythm had been established, collaborators did not perceive the need to adapt further, indicating that a state of effortless collaboration had been reached. Contrasting the quote from T1, other triads observed (e.g., T4, T5, T6) that the temporal synchronisation `just happened', implying the absence of conscious effort to reach a synchronised state. Still, synchronisation happened nonetheless, as expressed by e.g., T5: \textit{`I think once we got in the rhythm...Yeah. Then we were able to kind of just keep going.'}.

\subsection{Leader and Follower}\label{sec:LF}
The data analysis revealed two distinct types of leader-follower patterns: static 
and absent. Interestingly, this finding showed clear differences between the dyadic and triadic strategies employed. During most dyadic---7 out of 10 with the exception of D11, D14, and D16---collaboration, we could observe a strong pattern emerge focusing on a static leader---which was defined by the task at hand---namely the cuber. The static pattern is characterised by the absence of change on who is leading the collaboration. Most participants in the dyadic configurations perceived the cubers task as more difficult, making this task the bottleneck which limited the pace at which the group could perform. Given that this task was being perceived as slower, the bowler would typically be ready at the centre of the table before the cuber would be ready with the next cube. Therefore, the cuber was often perceived as the leader during the dyadic collaboration, setting the collaborative pace to be followed. As expressed by D15: \textit{`I like to believe that I set the speed [...] I don't recall waiting for the bowl very often'}. Seven of the ten dyadic groups reported similar observations, describing the slower sub-task performer to be leading. Contrasting these seven groups, two groups (D14 and D16) described the absence of a distinguishable pattern. They describe that each dyad member was seemingly independent of one another, completing their task. The dyads found a rhythm that led them to complete their sub-task at the same speed without needing a dedicated `leader'. This was expressed by, e.g., D14: \textit{`I don't think so. Okay, I think they seem pretty equal.'}.



Contrasting the dyads, the triads were less explicit about the presence of a leader. Given that each person interacted with two other collaborators instead of only one, the presence of two leaders was a possibility, as expressed by the bowler in T1. Here, the bowler followed both cuber's behaviour, who were leading the interaction and providing the rhythm to which the bowler adjusted their pace. Due to the synchronisation between collaborative efforts of the cubers, the two leaders, i.e., the cubers, were perceived as providing one unified collaborative rhythm by the bowler.

\begin{quote}
    \textit{`I feel I [the bowler] kind of am the follower for their collective behaviours, because I was trying to match the tempo. But also in this case, because I personally see the dropping of the cube as a cue, an audio cue on what the rhythm is.'} - T1 
\end{quote}

Finally, some groups described the absence of a dedicated leader. This was, for instance, expressed by one of the cubers in T8 who stated that: \textit{`I don't think there is a leader role, but I kind of follow her because I think she put the cube in first, and then I will follow.'} However, as evident from the quote, it can be argued that---even though the cuber does not call it a leader role---they are still following the other cuber.


\subsection{Interpersonal Communication}\label{sec:Communication}


To investigate differences in communication between dyads and triads, we used the recorded audio-video data. Each video was coded by hand to identify the occurrence of speech as well as the topic discussed. We counted each time a verbal interaction occurred. A new interaction was characterised by a change of topic from one category to another or after a conversation was started following more than five seconds of silence. As with the previous finding (see~\Cref{sec:LF}), we were able to observe difference between the two configurations. The distribution of communication, for dyads and triads, by different topics can be seen in~\Cref{fig:Pareto}.

\input{Chapters/Pareto}

While both dyads and triads conversed during task completion, we observed one key difference between communicative behaviour. As shown in the Pareto diagrams (\Cref{fig:Pareto}), the first two bars (i.e., `Small talk' and `General task related') are reversed in order between the dyads and triads. Specifically, the most frequent topic of communication for dyads was task unrelated `small talk' (33\%), contrasting only 19.7\% of small talk in the triadic setting (e.g., \textit{`Have you watched Gravity Falls [Netflix show]?'} - T5). This ordering was switched, as `general task-related' communication (e.g., \textit{`I think I just dropped one cube [on the floor]'} - D13) was the most common topic of conversation during the triadic task completion (40.7\%) and the second most frequent for the dyads (25.4\%). Results showed a significant difference among the two groups in terms of `Small talk' to `General task related' conversation ($\chi^2$(1, \textit{N}~=~20)~=~5.991, \textit{p}~=~.0144).

\subsection{Point-of-Assembly}\label{sec:PoA}
While entrainment primarily refers to the temporal synchronisation of actions, the spatial aspect is equally important. As a measure of this, we investigated two aspects of the collaboration. Firstly, the consistency of the point of assembly (\Cref{subsubsec:ConsistentPoA}), and secondly, the consistency of the participants hands trajectories and how this consistency---or the absence of it---evolved over time (\Cref{subsubsec:ConsistentTrajectories}).



\subsubsection{Consistency in Point-of-Assembly}\label{subsubsec:ConsistentPoA} We identified two different strategies utilised to negotiate the point-of-assembly (PoA). The PoA refers to the specific spatial location where the bowl is positioned, i.e., moved to or held, for the cube(s) to be dropped. Through examination of the video footage and further elaboration through the interviews, we discerned two primary strategies employed when approaching the PoA: 1) optimisation for group efficiency, and the 2) optimisation of individual task.


For strategy 1), we observed the tendency to prioritise adjustment of the individual task, while  not emphasising the own ease of task completion, but rather optimise for the other collaborators. This strategy was motivated by the intent to increase overall group efficiency by adjusting one's own task to facilitate one's collaborators. 
Several dyads and triads (e.g., T2, T4, T8, D12, D16) expressed this behaviour. The bowler in T4, for instance, used this strategy to accommodate the two cubers: \textit{`They [the cubers] were putting the cubes down at the same time. So I just need to place it in the middle [to make it easy for them].'}. A similar approach was described by the bowler in T2 who optimised the placement of the bowl to increase overall group efficiency. Even though this increased the range of motion needed by the bowler, it reduced the task complexity for the two cubers.

\begin{quote}
    \textit{`I also optimise the location of the bowl, and then we can move it further [away from the bin, i.e., from the cubers end-point] because on a global sense, that would be the best strategy considering both of your arms [the cubers]'} - T2 
\end{quote}






A different, albeit less frequent, approach was observed in which several participants of both roles reported on the optimisation of their own task---while de-emphasising the benefit to the other collaborators. This change in strategy, 
has the potential to create a ripple effect, influencing the behaviour of other participants and ultimately enhancing the overall efficiency of the group's tasks.

In, e.g., D11, the bowler describes reducing the distance between the bowl and bin to increase the ease of dropping the cubes. This adjustment, to make their own task easier, resulted in an increase in distance for the cuber, i.e., making the cubers task more demanding. Nevertheless, while the motivation was to ease one's own task, the collaborators described an overall efficiency increase, as the adjustment resulted in an optimisation of the bowler's task, which in this specific dyad was perceived to be the slower task:

\begin{figure*}[t]
    \centering
    \includegraphics[width=\linewidth]{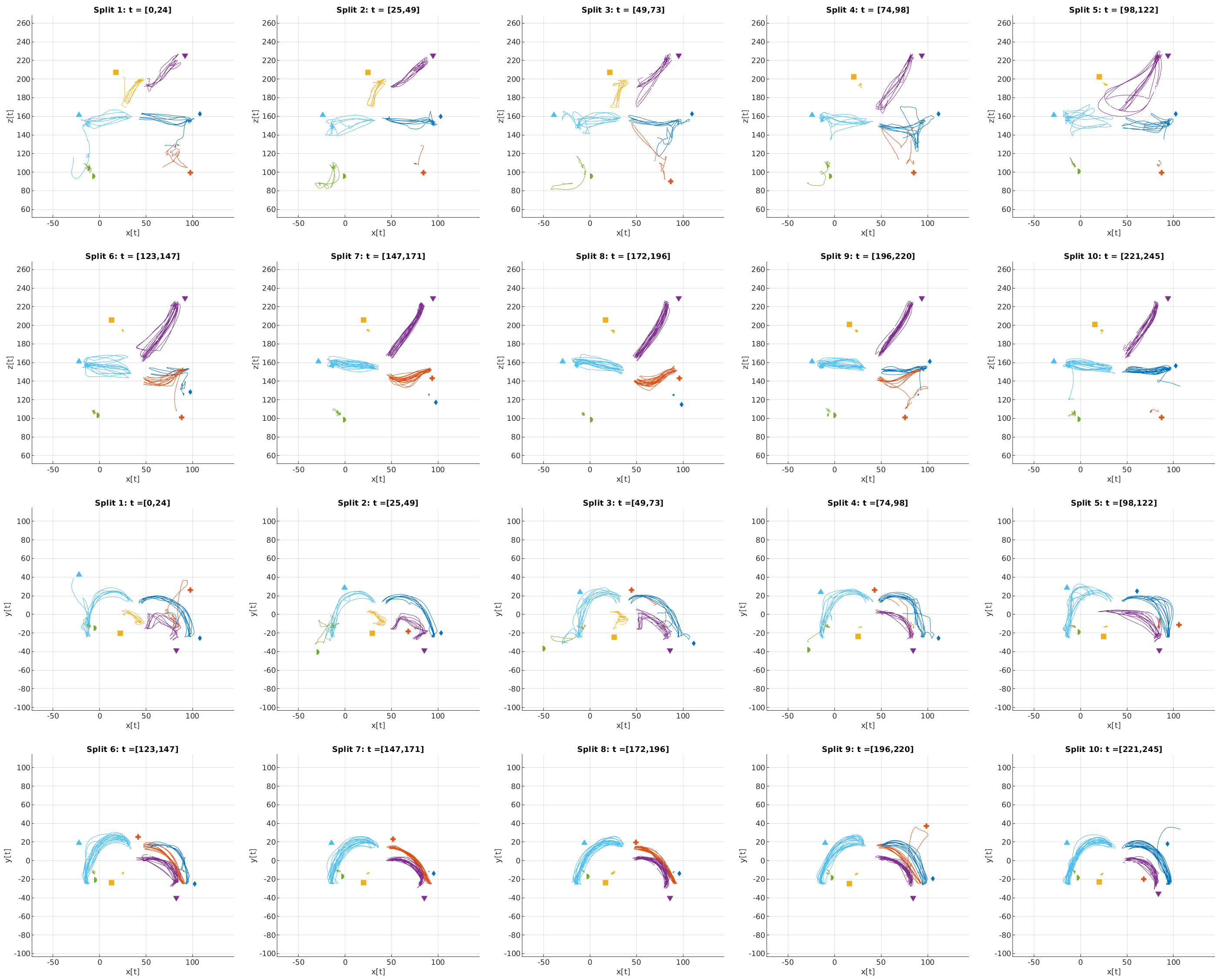}
    \caption{This figure shows 20 frames (5 columns, 4 rows). The first two rows represent the x-z perspective while the second two rows represent the x-y perspective (all for the same triad T3). Each frame presents six lines, two for each collaborator (right hand/left hand). Each frame plots the trajectory for each of the six collaborating hands for 10\% of the task duration (i.e., 24 seconds). As clearly visible, the consistency of trajectory increases which is indicative of spatial synchronisation. \textit{Colour coding:} Yellow (\mysym[brightYellow]{110}) and purple (\mysym[purple]{116}): left and right hand of the bowler. Light blue (\mysym[lightBlue]{115}) and green (\mysym[green]{119}): left and right hand of cuber one. Dark blue (\mysym[darkBlue]{169}) and orange (\mysym[orange]{58}): left and right hand of cuber two.}
    \label{fig:G3MoCap}
\end{figure*}
\begin{quote}
    \textit{`...[the cuber] should be the one who decides speed and I can adjust to it. [...] later, I found out, I'm actually the slower one. So then, I just moved the bowl near me so that it's easier for me to put [the cubes] in.'} - D11
\end{quote} 

Similar findings were observed in the triadic setting. In T5, for instance, the bowler reaches the same conclusion by adjusting the position of the bowl, aiming to minimise the distance to the bin, thereby making the task easier for themselves. This adjustment reduces the amount of movement required for their own task, however, while the motivation was the improvement of ones own task, this may or may not lead to a decrease in ease of task completion amongst the collaborator(s).

\begin{quote}
    \textit{`[The bowl] was more to the side [towards the bin], but I wouldn't say it was too far. It's still okay. [...] This might have been inclined towards the bin side, in this case, to my left side'} - T5 
\end{quote}

While groups using either strategy, performed these actions in order to optimise the efficiency, the difference was the targeted efficiency increase, i.e., the individual or the groups. Whereas the adjustment of own behaviour to facilitate collaborators prioritises an improvement to group efficiency, by reducing the collaborators workload, the second approach achieves this only as a potential side effect resulting from the individual optimisation. Regardless of the strategy utilised, both dyads and triads (e.g., T1, T3, T5, T7, D11, D12) highlighted the importance of consistency to achieve high task efficiency. This finding contrasts with the expected main contributor to efficiency---speed. Higher consistency led to greater predictability in motion, making it easier to anticipate the PoA and resulting in less downtime during task completion. This was, for instance, expressed by the bowler in T3, as described below. 

\begin{quote}
    \textit{`[...] I keep on doing this sweeping motion, their goal is to put it [the cube] in the bowl. But then the goal point keeps on moving. It might lead to more confusion. So I thought it'd be better [...] just keep it constant so that they can predict the next move.'} - T3 
\end{quote}
The here described attempt at using a sweeping motion for the bowl can be seen in Split 5 in~\Cref{fig:G3MoCap} (purple~\mysym[purple]{116}). However, after initial experimentation with a moving PoA, the group focused on a static PoA to increase predictability.

In addition to participant descriptions, the consistent PoA selection was evident from the video analysis. After, e.g., T4 had found a position that worked for them---the bowl slightly closer to the side of the bin---it was clear to see that an effort was put into keeping the PoA as consistent as possible. \Cref{fig:ConsistentPlacement} illustrates the bowl placement for T4 over the course of 60 iterations~$(24 \rightarrow 44 \rightarrow 64 \rightarrow 84)$.

\subsubsection{Consistency in trajectories}\label{subsubsec:ConsistentTrajectories}

In addition to the consistency of the PoA, we further investigated the consistency of trajectories and the development thereof over time. 
To illustrate this, we visualised the collected motion capture data and created plots of the trajectories from two different angles (x-y and y-z perspectives) in windows of 24 seconds each (leading to 10 windows for the 4 minute task completion). \Cref{fig:G3MoCap} presents one example for a triad (T3) illustrating the hand trajectories of the six hands of the three collaborators over the course of the four minutes. Similar trends were found in other dyads and triads. The division into 24 second windows was chosen as it allowed for effective visualisation of the task over time, leading to us being able to observe variations in consistency of participants trajectories throughout the course of the task completion.

From the presented time series, it became apparent that the behaviour of each member of the triad changed during the four-minute task completion. Initially, all three participants used both hands---albeit not necessarily to the same extent, as evident from the trajectories presented in~\Cref{fig:G3MoCap}. Each collaborator is represented with two colours, one for each hand. Specifically, cuber one: green and light blue ( \mysym[green]{119} / \mysym[lightBlue]{115} ), cuber two: orange and dark blue ( \mysym[orange]{58} / \mysym[darkBlue]{169} ), and the bowler: purple and yellow ( \mysym[purple]{116} / \mysym[brightYellow]{110} ). However, within the first two minutes of the task, all three participants transitioned to using only one hand, indicated by the disappearance of the green, orange, and yellow trajectories. However, cuber two shortly after switched hands, as seen in split 6 to 9, only to switch back to dark blue in split 9/10. Following split 5, both the starting and ending points of the trajectories, as well as the paths taken, became increasingly consistent. These trajectories are indicative of each participant finding a working rhythm for their part of the task.

\subsection{Multisensory Information}\label{sec:Sensory}
An additional characteristic identified through the study, and highlighted by both dyads and triads, is related to the dependence on, at times non-obvious, sensory information. While the task described required tactile and visual information, i.e., feeling the cubes and seeing where to drop them, the importance of auditory information was observed during the video analysis and described by participants in the post-task interviews.



While the use of visual and tactile information was expected, most groups (e.g., T1, T5, T8, T9, D12, D17) further relied on auditory cues produced by the task. 
Here, the benefit of the auditory signalling was expressed in multiple directions, specifically both to and from the bowlers' task. As it could at times be difficult to see if the small cubes were dropped by the cuber(s), several bowlers reported using sound to confirm that the cubes were in fact in the bowl. In addition to the visual cue of seeing the cuber(s) hand(s) over the bowl, the auditory cue further provided information about when the bowl could be moved towards the bin, thereby freeing visual cues as the conveying information did not, exclusively, rely on visual inspection. Audio cues were relevant in both the dyadic as well as the triadic task completion.

\begin{quote}
    \textit{´[...] also auditory. Yeah, listening. Yes. Because I would look, but I also make sure that I heard something.'} - T8\\\\
    \textit{`When she [the cuber] dropped the cube into the bowl, that's when I knew it's ready to dump it [into the bin].'} - D20
\end{quote}

During the video inspection, we could observe that bowlers stopped the interaction when they missed the bin during the emptying of the bowl. This was due to the absence of audio cue, as the cubes did not hit the bin, but landed on the carpet (see supplemental video material $\sim$ 00:07--00:13 (dyadic) and 01:17 (triadic), not producing the distinct sound the bowler was listening for as confirmation of successful task completion.


Just as the bowler used audio cues to know when to proceed to the next step of their task, so did the cuber(s). In the cuber(s) case the acoustic signal used was caused by the bowler dropping cubes in the bin next to them, thereby signalling the end of the iteration. This sound of falling cubes would signal the cuber(s) that the bowl is about to return to the PoA to collect the next cube(s). 

\begin{quote}
    \textit{`Sound. Definitely! I think, whether I knew it or not, I think I was cue'ing into that sound of \textbf{dum dum} [sound of cubes dropping into the bin].'} - T5
\end{quote}

These findings highlight the value of multi-modal signalling during task completion. The use of, e.g., acoustics, frees other senses for the preparation of the next iteration, while simultaneously providing feedback about the progression of the iteration.

%% file: Chapters/Pareto.tex
\begin{figure*}[h]
    \centering 
    \begin{subfigure}[t]{0.48\textwidth}
        \centering
        \includegraphics[width=\textwidth]{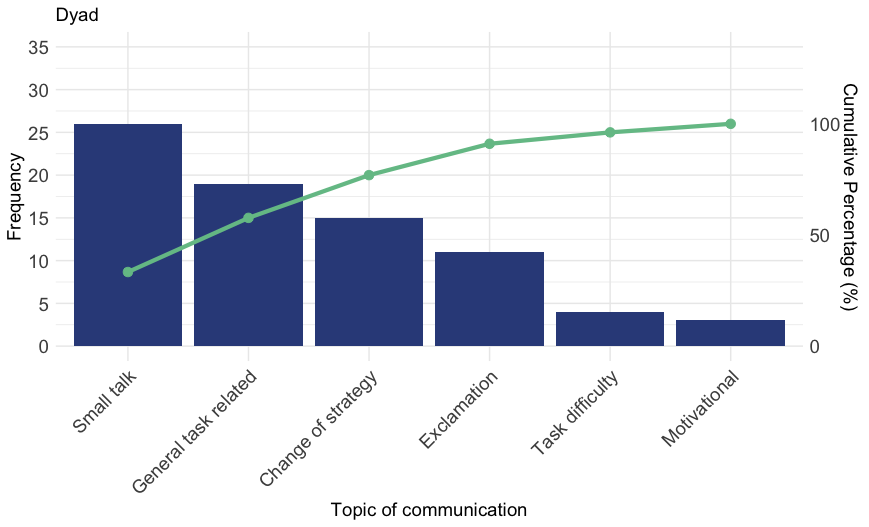}
    \end{subfigure}%
    \hspace{3mm}
    \begin{subfigure}[t]{0.48\textwidth}
        \centering
        \includegraphics[width=\textwidth]{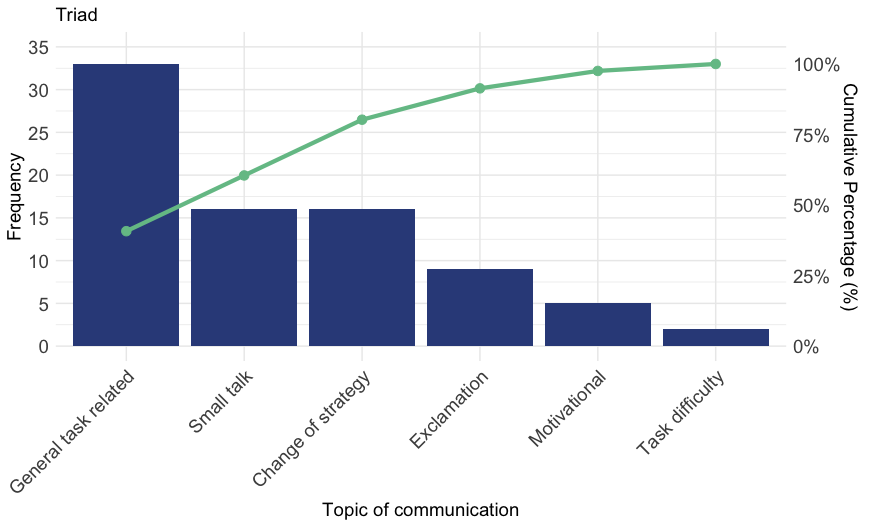}
    \end{subfigure}
    \caption{Pareto diagrams for the dyads (left) and triads (right) respectively. The data visualised is the topic of communication. The graphs show, that while both dyads and triads discussed the same topic categories, the first two columns `Small talk' and `General task related talk' were inverted. This means that the dyads focused on task unrelated conversation without a decrease in task performance (average iterations pr. 10 sec interval, see~\Cref{tab:DyadTriad}) compared to the triads who focused conversation on the task at hand. This could be indicative of the dyadic collaboration requiring less mental workload.}
    \label{fig:Pareto}
\end{figure*}

%% file: Chapters/Discussion.tex
\section{Discussion}
In this paper we have presented a mixed-method study focusing on entrainment during task completion in human pairs and groups. This paper contributes to the HCI and HRI literature, by investigating characteristics of entrainment during dyadic and non-dyadic collaboration. 
This section will discuss some of these findings in relation to existing literature and present three design considerations for the design and improvement of human-robot collaboration (HRC) using collaborative robots (cobots).

\subsection{Differences in Interpersonal Communication}\label{subsec:Workload}
As highlighted in~\Cref{sec:Communication}, our study showed that participants in both conditions, i.e., dyads and triads, communicated while performing the collaborative task. We categorised this communication into several different topics such as `Small talk', `General task related' or `Change of Strategy', and while we could observe that both conditions talked about the same topics (i.e., categories), differences in frequencies were observed.
Here it is particularly interesting to note that the dyadic condition had a significantly higher amount of off-topic conversation (e.g., movies or weekend plans) over task-related conversation compared to the triads (33\% compared to 19.7\%). While we did not collect evidence about perceived task complexity, we have presented evidence (see~\Cref{sec:Performance}) that indicates that no significant differences in performance was identified between conditions. 
The consistency in average iterations between dyads and triads is indicative of a natural, or comfortable, frequency---regardless of the group formation. While the performance is the same, the degree of off-topic conversation increased by 13.3 percentage points in the dyadic configuration. 

A potential hypothesis leading to this could therefore be related to a lower mental workload required for the task in the dyadic configuration compared to group-based collaborations. This would allow a higher degree of non task related conversations without impacting the task performance. To investigate this, and be able to confirm or refute this hypothesis, a follow-up study utilising subjective instruments such as the NASA-TLX~\cite{NasaTLX} or the (simplified) subjective workload assessment technique~\cite{SSwat}, performance based mechanisms, such as the tone detection task~\cite{butterfield2006correction}, or strategies relying on physiological data such as measured through EEG or fNIRS~\cite{Aghajani:2017:EEGFNIR} are needed. 

An alternative explanation to the change in communication patterns might be related to the \textit{many minds problem} as presented by~\citet{COONEY202022}. In its essence, the authors argue that the move from a dyadic configuration to a group based setting, such as a triad, changes the way in which we communicate---and about what---considerably. In the here presented study, the observations presented in relation to the type of communication (see~\Cref{sec:Communication}) might be related to the many minds problem, i.e., when people interact in larger groups, they are less likely to disclose personal or private information and opinions (i.e., engage beyond the task at hand) as they are in dyadic configurations.

\subsection{Robot Adaptability}
Typical industrial robots, such as those used in the automobile industry, are characterised by their ability to perform very effectively given constrained repetitive tasks that require no adaptation and minimal human intervention. However, with the increase in un-caged collaborative robots, humans and robots are beginning to collaborate in close proximity (e.g.~\cite{Sauppe_2015_Baxter,Cheon:2022:Bounded,Cheon:2022:PlaceSocial}). While previous research has shown that humans can entrain on robots~\cite{Lorenz:2011,Rea:2019:Entrainment}, we argue that this can be improved upon by giving robots a higher sense of awareness of their collaborators. This would allow the robotic collaborator to adjust to the humans' pace. 
This robots awareness of the human collaborators performance, has the potential to led to a higher degree of adaptation from robot to human(s), ultimately strengthening the collaboration as it becomes bidirectional entrainment, instead of unidirectional. While this would be novel for collaborative robots, similar effects have already been documented for interaction with social robots~\cite{Ansermin:2017:Entrainment}.
We argue that, as demonstrated in the human collaboration presented in this paper, providing collaborative robots with a better sense of awareness regarding elements such as collaborator's movement speed, performance, or deviations in the PoA, this could enhance the robots' ability to adapt, and in turn collaborate, with in mixed human-robot configurations. This possibility of bidirectional adaptation ($Humans \leftrightarrow Robots$), contrasting humans one sided adaption to a non-adaptive robot, can help to increase the level of collaboration~\cite{Christiernin:2017:Collaboration} between humans and robots.

As presented in this paper, participants experienced the occurrence of synchronisation. While variations between strategies for reaching synchronisation (described as  `groove' or `rhythm') occurred, the motion capture data in conjunction with the collaborator's described experiences are indicative of this. From this, we derive our first design consideration for future research focusing on collaboration between cobot(s) and human(s).

\begin{quote}
    \textbf{Design Consideration 1:} \textit{Designers should consider how collaborative robots can adapt their behaviour based on human collaborators' fluctuations in performance, speed, and PoA deviations.}
\end{quote}


\subsection{Noise is not always Noise}\label{subsec:Noise}
Within industrial tasks such as production, manufacturing, or warehousing, noise is typically perceived as an undesirable presence having potentially harmful implications on workers' health~\cite{Rabinowitz:2011:Hearing}. Especially given the nature of this task, which was inspired by an industrial pick-and-placing task, one would typically be interested in reducing noise. Therefore, it stands to reason that the reduction of noise would be a desirable goal. In this study we demonstrated, that subtle acoustic signals---specifically noise related to the task at hand (i.e., the sound of the dropping cubes)---allowed the collaborator's visual attention to be directed at other parts of the task while still receiving feedback on the progress of the collaborative effort. 

In this study, we investigated mutual entrainment~\cite{Rinott:2021}, meaning we did not provide any external stimuli, such as a beat, to which to entrain. However, the pick-and-placing task produced intrinsic acoustic feedback on multiple steps. Specifically, the cube(s) dropping into the bowl and from the bowl into the bin. As presented in~\Cref{sec:Sensory}, this auditory cue provided a rhythm to which collaborators structured their task. Specifically, the auditory cue informed the bowler that the bowl can be moved to the bin as they hear the cubes land in the bowl, and the cubers know that the next iteration has started once they  hear the cubes drop in the bin. As not every task produces task intrinsic auditory cues, the addition of task-relevant acoustic signals could aid the collaborators, humans and robots, by giving additional stimuli to which to entrain to. Previous research has shown that auditory signals can be perceived as useful during robot approach behaviour~\cite{Lohse:2013:RobotNoise}, industrial collaboration~\cite{Sauppe_2015_Baxter}, or to increase positive sentiment toward robots~\cite{Thiessen:2019:Chair}. In this paper, we argue that task-relevant sounds during human-robot collaboration might provide additional stimuli for entrainment during collaboration. Based on these findings, we derived our second design consideration for collaboration with cobots.

\begin{quote}
    \textbf{Design Consideration 2:} \textit{Designers should consider how collaborative robots can utilise sensory channels to provide feedback emphasising key events during the collaboration.}
\end{quote}


\subsection{Short-term Consistency is Key}
A frequent observation was the emphasis on consistency leading to predictability. The importance of legibility in human and robot motion, i.e., the ease of reading and inferring the goal based on observed motion, to identify the robot's intent, has been emphasised in previous research~\cite{Dragan:2013:Predictability}. Based on our findings, we argue that collaborative robots need to be consistent within the short-term, i.e., between individual iterations,  while allowing for long-term adaptations, allowing for change in performance over the course of several iterations. Naturally, what constitutes `short-term' and `long-term' varies greatly depending on the task. In the here presented task collaborators were able to perform each iteration within 2.27 seconds while other tasks (e.g., the assembly of several sub-components) might take significantly longer to complete. This is particularly important in tasks in which the robot initiates the task, i.e., is defining the pace at which the collaborations occurs. We, therefore, argue that the robot's movement and approach to solving a given task should not only be legibly---but that the need for legible behaviour can be reduced through short-term consistency as observed during human-human task completion (see~\Cref{sec:PoA}).
Based on this, we pose our third design consideration, emphasising the need for consistency.

\begin{quote}
    \textbf{Design Consideration 3:} \textit{Designers should consider how collaborative robots can exhibit short-term consistency, allowing them to be consistent between iterations, while also allowing for long-term adjustments based on the human collaborators behaviour.}
\end{quote}







\subsection{Limitations and Future Work}

As this lab study was conducted in a controlled environment, no external noise was present. However, the task chosen, resembling an industrial pick-and-placing task, is typically performed in environments with high potential for machinery producing noise. Therefore, the value and desirability of auditory cues, as presented in~\Cref{sec:Sensory} and used as the foundation for the second design consideration (see~\Cref{subsec:Noise}), might be affected by the presence of other auditory information. To identify the usefulness of auditory information in noisy environments, a follow-up study focusing on ecological validity needs to be conducted. Furthermore, as this was a novel task for the participants, different observations might occur when investigating this in a longitudinal context. 

Future work includes implementing and evaluating the proposed design considerations to assess their effect on human-robot entrainment in different configurations (e.g., dyads and triads). A follow-up study will have the potential to highlight the usefulness of the proposed design considerations in order to achieve synchronisation amongst humans and robots. This could be of particular interest in collaborative settings where the maximum working speed between humans and robots is not well aligned. Furthermore, this could provide insights into the side effects of human-robot entrainment during industry-inspired, fast-paced, collaborative tasks in group configurations dyadic setting investigated in current research. An additional future research direction could involve comparing cognitive workload during the collaboration between dyads, triads, and larger groups (as described in~\Cref{subsec:Workload}). Additionally, a quantification of the here presented findings identifying exactly when entrainment has occurred might make the findings presented in this paper more actionable.
Lastly, the verification of the findings presented in this paper through the investigation of different tasks could be beneficial. The authors of this paper are currently investigating the data from a follow up study using the `envelope stamping' task, which was piloted and mentioned in~\Cref{sec:taskDev}.




%% file: Chapters/Conclusion.tex
\section{Conclusion}

In this paper, we present the findings of a mixed-method laboratory study that investigates how human dyads and triads synchronise with each other temporally. To achieve this, we designed a fast-paced, short-cycle repetitive task inspired by an industrial pick-and-place scenario. We collected empirical data through a mixed-method approach, which involved interviews, audio-video recordings, and motion tracking of collaborators' hands and objects of interest.

Overall, we observed strong spatial consistency within participant groups, especially at the point of assembly, along with minimal temporal fluctuations in task performance, indicating the occurrence of entrainment. Specifically, we outline five key characteristics of how human dyads and triads entrain with each other. These characteristics relate to the occurrence of synchronisation, variations in leader-follower dynamics, distinctions in communication between dyads and triads, the significance of the point of assembly, and the impact of unintended noise generated by the task. Finally, we discuss three design considerations that will inform future research in the field of human-robot collaboration.